\documentclass{PoS}
 \usepackage{epsfig}
\usepackage{wrapfig}

\title{New Results from the MINOS Experiment}

\ShortTitle{New Results from the MINOS Experiment}

\author{\speaker{Anna HOLIN}%
        \thanks{On behalf of the MINOS Collaboration}\\
       University College London\\
       E-mail: \email{annah@hep.ucl.ac.uk}}

\abstract{The MINOS experiment is a long-baseline neutrino experiment designed to study neutrino behaviour, in particular the phenomenon of neutrino oscillations. MINOS sends the NuMI neutrino beam through two detectors, a Near Detector 1 km downstream from the beam source at Fermilab, and a Far Detector 735 km away in the Soudan Mine in Minnesota. MINOS has been taking beam data since 2005. This document summarises recent neutrino oscillations results, with particular emphasis on electron neutrino appearance, which probes the angle $\theta_{13}$ of the neutrino mass mixing matrix. For an exposure of 8.2$\times 10^{20}$ protons on target, MINOS finds that $\sin^{2}(2\theta_{13})<0.12$ for the normal mass hierarchy, and $<0.20$ for the inverted mass hierarchy at the 90\% C.L., if the CP-violating phase $\delta=0$.}

\FullConference{XXIst International Europhysics Conference on High Energy Physics\\
		 21-27 July 2011\\
		 Grenoble, Rhones Alpes France}

\begin{document}

\section{Neutrino Oscillations}
In the past 20 years, one of the most significant developments in particle physics was the discovery of neutrino oscillations \cite{ref0}-\cite{ref5}, indicating that neutrinos have mass. There are three generations (flavours) of neutrino, the electron, muon, and tau neutrino ($\nu_{e}$,$\nu_{\mu}$,$\nu_{\tau}$). Each neutrino also has a corresponding anti-particle. Neutrino oscillations are parametrised using two mass squared differences ($\Delta m_{21}^{2}$ - solar sector, and $\Delta m_{32}^{2}$ - atmospheric sector), three mixing angles ($\theta_{12}$, $\theta_{13}$, and $\theta_{23}$), and a CP violating phase ($\delta$). 

\section{The MINOS Experiment}
The MINOS experiment \cite{ccprd} is designed to probe neutrino behaviour and the phenomenon of neutrino oscillations by sending the NuMI neutrino beam through two detectors \cite{detec}, the Near Detector at Fermilab, 1km downstream of the beam source, and the Far Detector in the Soudan Mine in Minnesota, 735km away. Both detectors are magnetized iron scintillator tracking calorimeters and are designed to be functionally identical to allow cancellation of certain systematic errors, like for example any mismodelling of the neutrino flux or cross-section. The NuMI neutrino beam is created by impacting 120 GeV protons onto a thin graphite target. The resulting hadrons (mostly pions and kaons) are collimated by two magnetic horns and then decay producing a beam of mostly $\nu_{\mu}$ with a small 7\% component of $\overline{\nu_{\mu}}$, and a 1.3\% contamination of beam $\nu_{e}$ and $\overline{\nu_{e}}$. It is possible to reverse the horn current so as to achieve a beam with a higher proportion of muon antineutrinos.

\section{Muon Neutrino Disappearance}
MINOS oscillation analyses use the Near Detector to measure the neutrino interaction rate before any oscillations have occurred. These measured data are then extrapolated to the Far Detector to predict what would be seen in the absence of neutrino oscillations. This final data set is then unblinded and compared to the prediction. In the case of the $\nu_{\mu}$ disappearance analysis, the survival probability of a muon neutrino is given by:
\begin{equation}
P(\nu_{\mu}\rightarrow \nu_{\mu})\approx 1-\sin^{2}(2\theta_{23})\sin^{2}(1.27\Delta m^{2}_{32} (L/E))
\end{equation}

\begin{figure}
\centering
\begin{tabular}{cc}
\epsfig{file=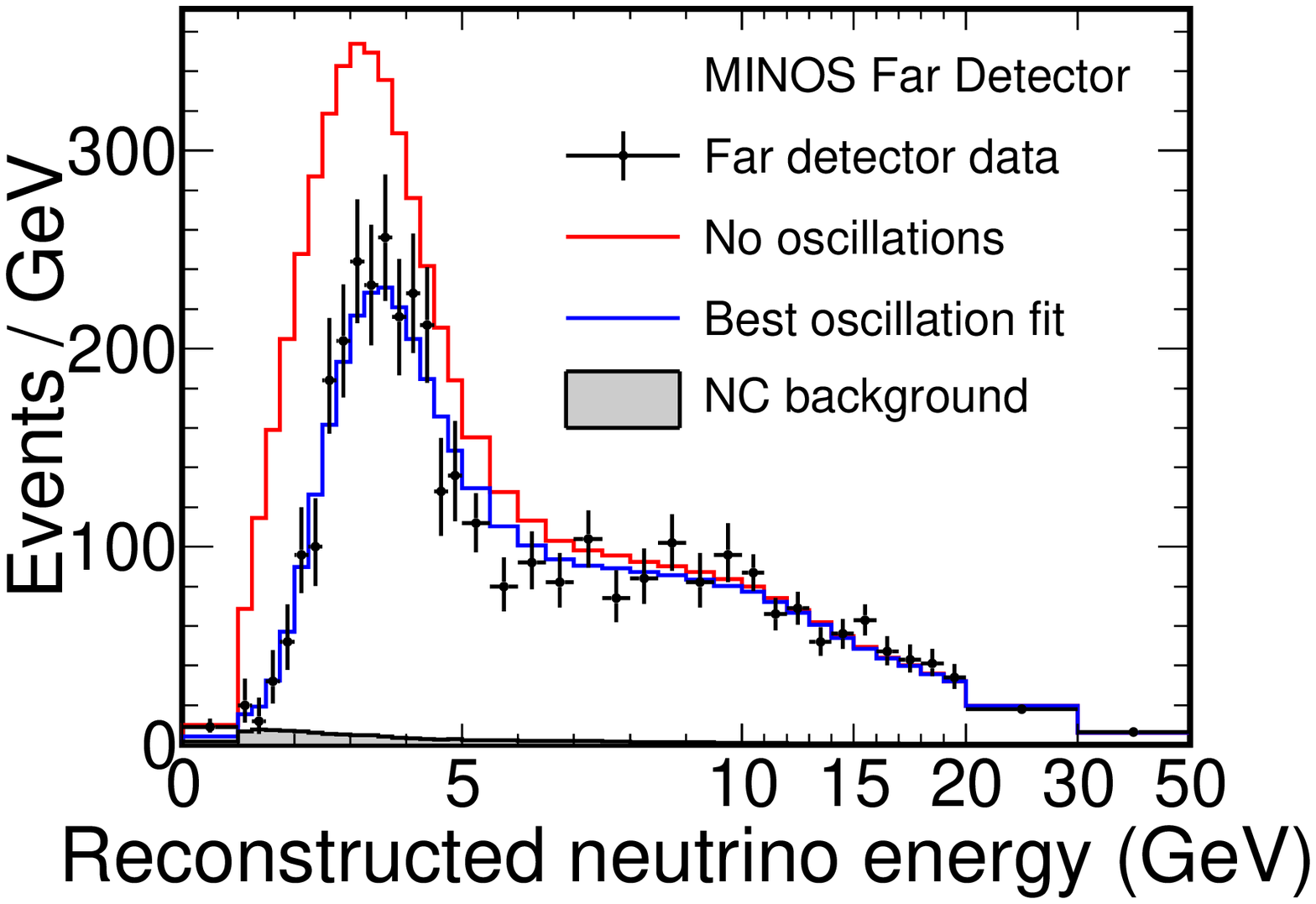,width=0.45\linewidth,clip=} & 
\epsfig{file=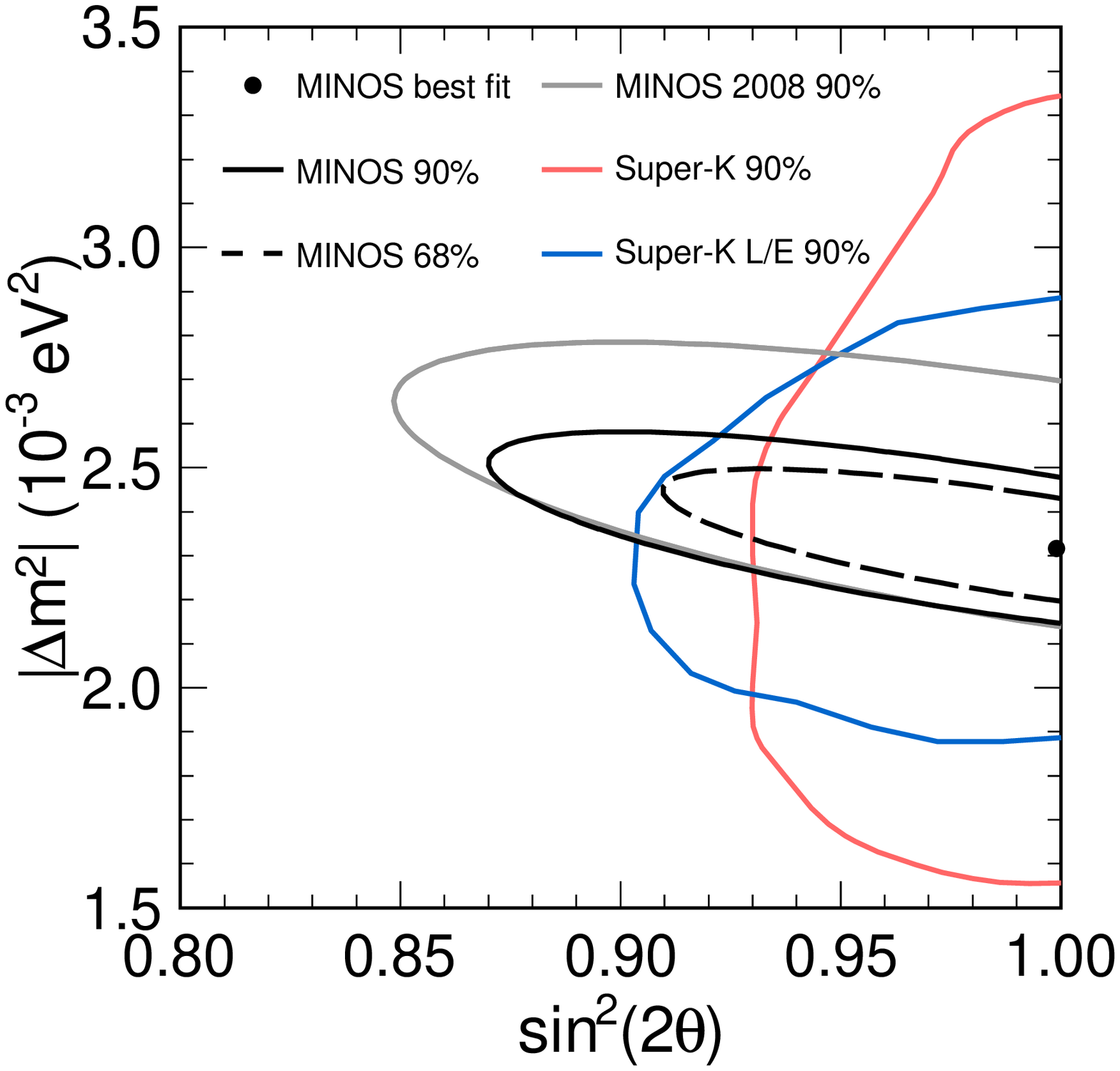,width=0.45\linewidth,clip=} \\
\end{tabular}
\caption{The left plot shows the energy spectrum of fully reconstructed Far Detector events classified as charged current interactions in black. The red histogram represents the spectrum predicted from measurements in the Near Detector assuming no oscillations, while the blue histogram reflects the best fit of the oscillation hypothesis. The shaded area shows the predicted neutral current background. The right plot shows the likelihood contours of 68\% and 90\% C.L. around the best fit values for the mass splitting and mixing angle. Also shown are contours from previous measurements \cite{ref1,ref2}.}\label{fig:numuspectrum}
\end{figure}

Figure  \ref{fig:numuspectrum} shows the results of the $\nu_{\mu}$ disappearance analysis for an exposure of 7.25$\times 10^{20}$ protons-on-target (POT). For this data set, 2451 charged current muon neutrino events were predicted in the Far Detector fiducial volume, but 1986 events were observed \cite{ccprl}. Consequently, the atmospheric mass-squared difference was found to be $\Delta m^{2}=2.32^{+0.12}_{-0.08} \times 10^{-3}$eV$^{2}$, and the mixing angle parameter $\sin^{2}(2\theta)>0.90$ (90\% C.L.). Exotic neutrino models like neutrino decoherence and neutrino decay were excluded at the 9$\sigma$ and 7$\sigma$ level respectively.

\section{Muon Antineutrino Disappearance}
MINOS has taken some data in antineutrino (reversed horn current) mode and carried out an antineutrino disappearance analysis \cite{numubarprl}. For an exposure of 1.71$\times 10^{20}$ POT, MINOS predicted 156 charged current $\overline{\nu_{\mu}}$ events in the absence of oscillations, however, 97 events were observed, thus disfavouring the no oscillations hypothesis at the 6.3 $\sigma$ level. The best fit antineutrino oscillation parameters were found to be $|\Delta \overline{m}^{2}|=(3.36^{+0.46}_{-0.40}$(stat.)$\pm0.06$(syst.)$)\times 10^{-3}$ eV$^{2}$ and $\sin^{2}(2\overline{\theta})=0.86^{+0.11}_{-0.12}$(stat.)$\pm0.01$(syst.).

\section{Electron Neutrino Appearance Analysis}
Muon neutrinos may oscillate into electron neutrinos as they travel from the Near to the Far Detector. The corresponding oscillation probability is to first order given by:

\begin{equation}
P(\nu_{\mu}\rightarrow \nu_{e})\approx \sin^{2}(2\theta_{13})\sin^{2}\theta_{23}\sin^{2}(1.27\Delta m^{2}_{atm} (L/E))
\end{equation}

If the oscillation angle $\theta_{13}$ is non-zero, then this should manifest itself as $\nu_{e}$ appearance at the Far Detector. In the case of MINOS, the search for $\nu_{e}$ appearance \cite{nueprl} is however made very difficult by low statistics and by a large background of neutral current events that mimic charged current $\nu_{e}$ events. To disentangle any potential signal from the large backgrounds, a number of cuts are applied to the data. First, a number of data quality cuts like timing and fiducial cuts are applied to select good beam events. Then, since $\nu_{e}$-like events consist of electromagnetic showers, any events with long muon tracks are removed from the sample. Only events with at least one shower and a well-defined shower core, and within an energy range of 1-8 GeV, are selected for the final sample. Finally a selection algorithm is used to obtain the final data sample. This selection algorithm uses a MC library of 20 million signal and 30 million neutral current events to find the 50 best matches for each event and to construct three variables that are combined into a neural network to obtain a final discriminant variable (LEM). 

In order to use the Near Detector data to predict the Far Detector data, the former is decomposed into its components: 60\% neutral current events, 29\% short-track charged current $\nu_{\mu}$ events, and 11\% intrinsic beam charged current $\nu_{e}$ events after all selection cuts. Those components are then extrapolated to the Far Detector using Monte Carlo Far/Near ratios accounting for various systematic errors such as flux, cross-section, fiducial volume, energy smearing, detector effects, and muon neutrino oscillation parameters. A $\nu_{\tau}$  appearance component resultant from oscillated $\nu_{\mu}$ is also added to the Far Detector prediction.

\begin{figure}[hb]
\centering
\begin{tabular}{cc}
\epsfig{file=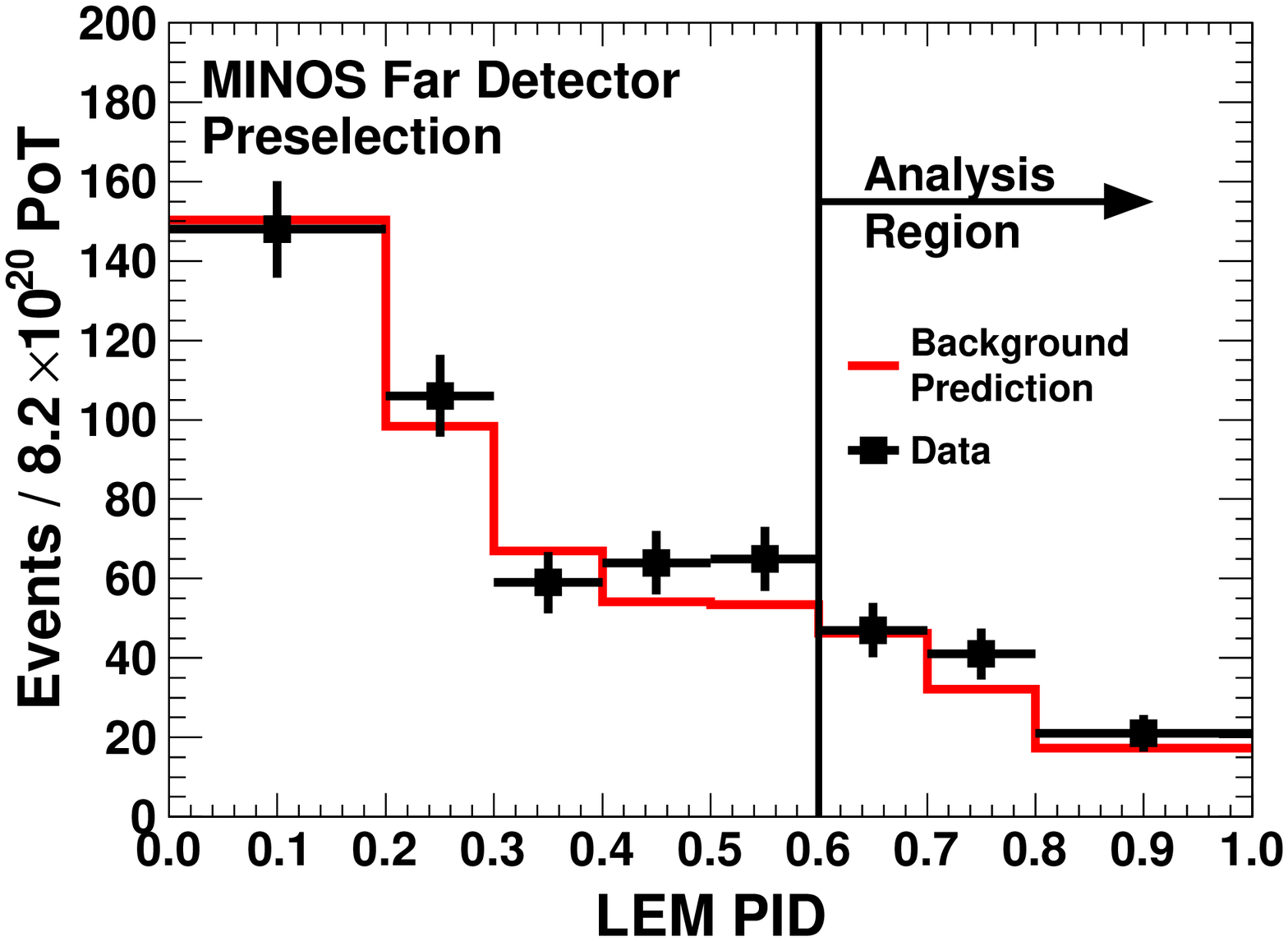,width=0.4\linewidth,clip=} & 
\epsfig{file=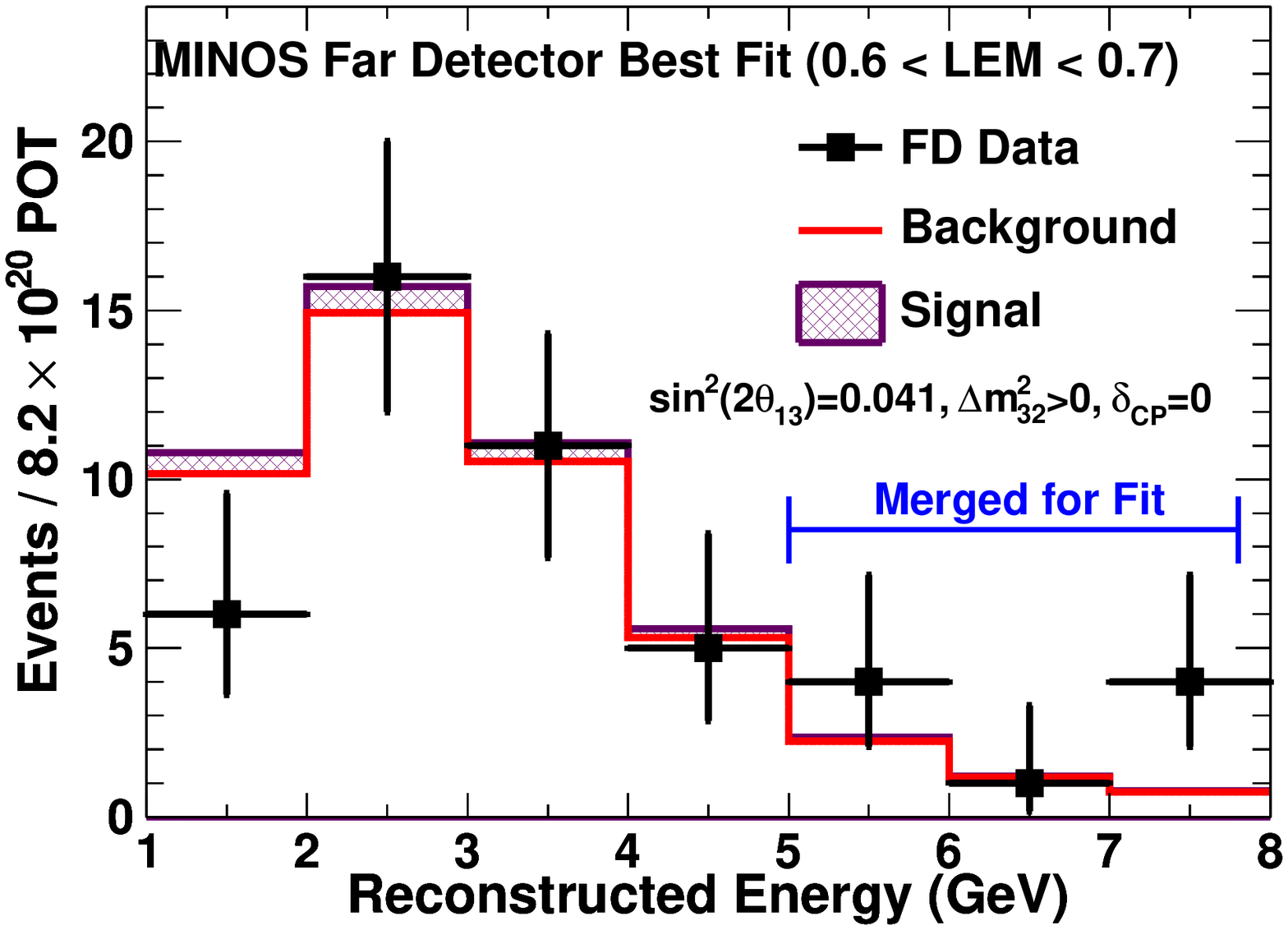,width=0.4\linewidth,clip=} \\
\epsfig{file=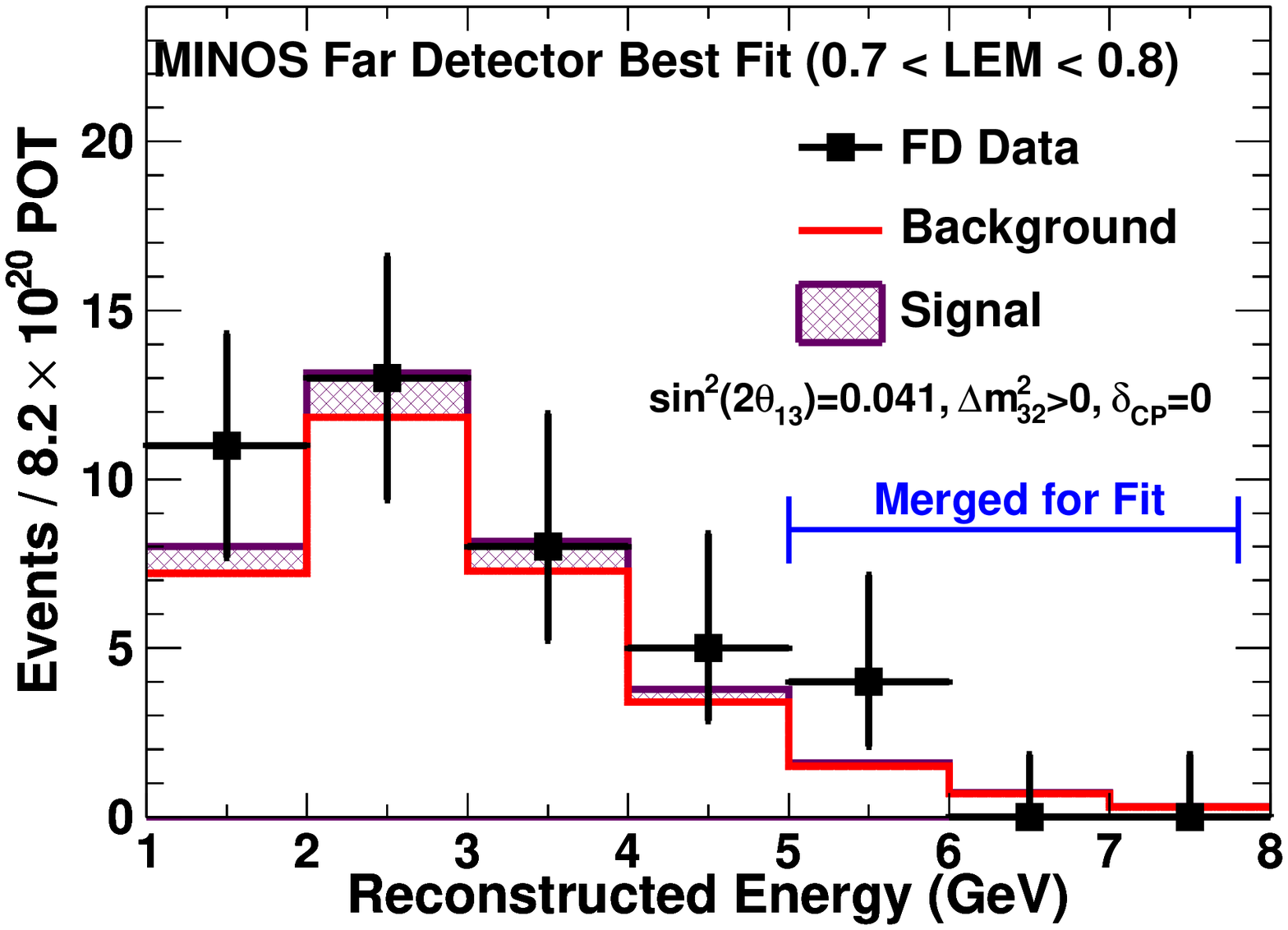,width=0.4\linewidth,clip=} &
\epsfig{file=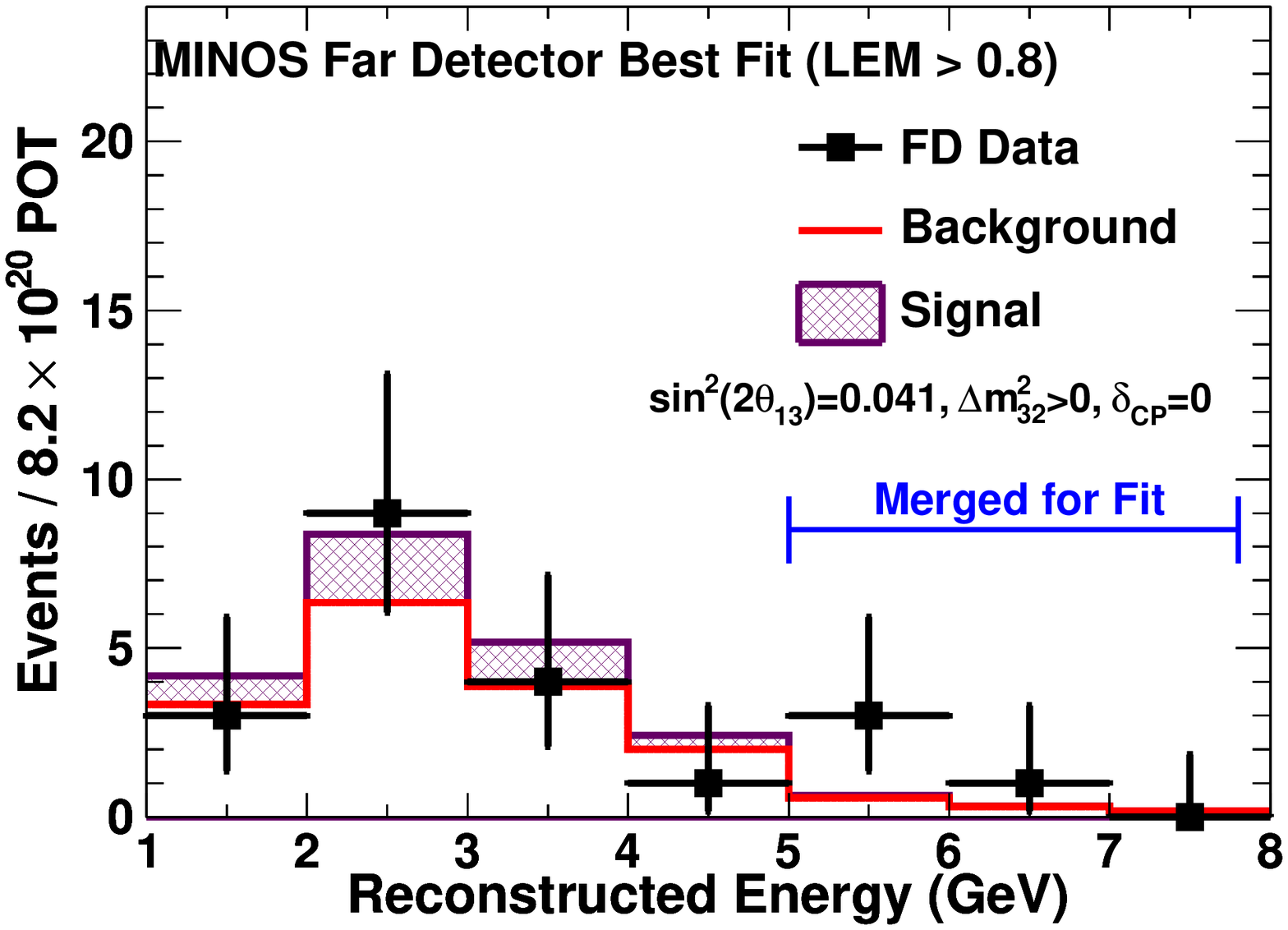,width=0.4\linewidth,clip=} \\
\end{tabular}
\caption{The top left plot shows the LEM discriminant variable distribution in the Far Detector. The top right plot and the bottom plots show the reconstructed energy spectra for charged current $\nu_{e}$ candidate events for the three LEM analysis bins. The black points show the data with statistical error bars. The red histograms show the expected background together with the contribution of $\nu_{e}$ appearance signal (hatched area) for the best-fit value of $\sin^{2}2\theta_{13}$ = 0.041.}\label{figspectra}
\end{figure}

\begin{wrapfigure}{l}{60mm}
  \begin{centering}
    \includegraphics[width=.42\textwidth]{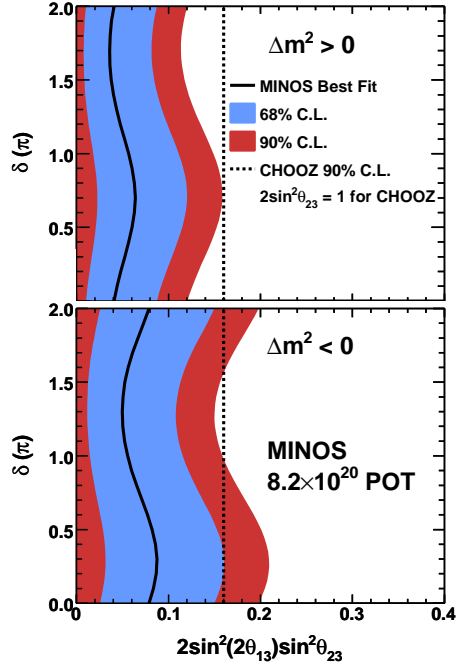}
\caption{Allowed ranges and best fits for $2\sin^{2}(2\theta_{13})\sin^{2}\theta_{23}$ as a function of CP-violating phase $\delta$. The upper panel assumes the normal neutrino mass hierarchy, and the lower panel assumes the inverted mass hierarchy. The vertical dashed line shows the Chooz 90\% C.L. upper limit assuming $\theta_{23}$=45$^{\circ}$ and $\Delta m^{2}_{32}$=2.32 eV$^{2}$.}
\label{figresult}
\end{centering}
\end{wrapfigure}

For the final analysis, MINOS uses five bins in reconstructed energy and three bins for the LEM discriminant variable (0.6-0.7, 0.7-0.8, and above 0.8). Systematic uncertainties taken into account include the composition of the Near Detector spectrum, the calibration (relative energy calibration, gains, absolute energy calibration), the relative Near/Far normalization, the hadronization model and other smaller uncertainties. In a signal enhanced region where LEM$>0.7$, MINOS predicts 49.6$\pm$7.0(stat.)$\pm$2.7(syst.) events in the Far Detector for an exposure of 8.2$\times 10^{20}$, and observes 62 events. The final Far Detector spectra can be seen in Figure \ref{figspectra}. If the final data are fitted to a neutrino appearance oscillation hypothesis, for a CP-violating phase $\delta$=0, MINOS finds $\sin^{2}2\theta_{13}<0.12$ for the normal mass hierarchy, and $\sin^{2}2\theta_{13}<0.20$ for the inverted mass hierarchy at the 90\% C.L.. The corresponding best fit values are $\sin^{2}2\theta_{13}$=0.041 for the normal mass hierarchy, and $\sin^{2}2\theta_{13}$=0.079 for the inverted mass hierarchy. Figure \ref{figresult} shows the final $\nu_{e}$ appearance limits obtained by MINOS. It can be seen that MINOS was able to exclude parameter space below the limit set by the Chooz experiment \cite{chooz} for all values of $\delta$ for the normal mass hierarchy.

\newpage
\hspace{60mm} \section{Conclusions}
The MINOS experiment has carried out a number of analyses that are helping to measure and hone in on the values of the neutrino oscillation parameters. In the atmospheric sector, MINOS has been able to carry out the world's most precise measurement of the oscillation parameter $\Delta m^{2}_{32}$. For the electron neutrino appearance analysis, MINOS has been able to set the world's tightest limit on $2\sin^{2}(2\theta_{13})\sin^{2}\theta_{23}$ for the normal mass hierarchy.

\vspace{10 pt}
\footnotesize This work was supported by the U.S. DOE; the U.K. STFC; the U.S. NSF; the State and University of Minnesota; the University of Athens, Greece; and Brazil's FAPESP, CNPq, and CAPES. We are grateful to the Minnesota Department of Natural Resources, the crew of the Soudan Underground Laboratory, and the staff of Fermilab for their contributions to this effort.

\end{document}